\title{Towards a BRICS Optical Transient Network (BRICS-OTN)}
    \author{David A.H. Buckley \& Vanessa A. McBride\\
	South African Astronomical Observatory (SAAO) \\
	PO Box 9, Observatory 7935, Cape Town, South Africa \\
		\and
	Ulisses Barres de Almeida \\
	Brazilian Center for Physics Research (CBPF) \\
        Rua Dr. Xavier Sigaud 150, Rio de Janeiro 22290-180, Brazil\\
        \and
	Boris Shustov            \\
	Institute of Astronomy of the Russian Academy of Sciences (INASAN)
	\\
    119017, Pyatnitskaya str., 48 , Moscow,  Russia \\
	\and
	Alexei Pozanenko \& Alexander Lutovinov \\
	Space Research Institute of the Russian Academy of Sciences (IKI), \\
	84/32 Profsoyuznaya Str, Moscow, Russia
	\and
	Amitesh Omar \\
	Aryabhatta Research Institute of Observational Sciences (ARIES) \\
    Manora Peak, Nainital-263001 \\
    Uttarakhand, India \\
	\and
	Jayant Murthy \& Margarita Safonova \\
	Indian Institute of Astrophysics (IIA) \\
II Block, Koramangala, Bangalore 560 034, India\\
	\and
	Liu Jifeng \& Roberto Soria \\
	National Astronomical Observatories, CAS (NAOC) \\
20A Datun Road, Chaoyang District, Beijing, China \\
}
\date{28 September 2020}
\documentclass[12pt]{article}
\usepackage{graphicx}
\usepackage{eurosym}
\usepackage{array, booktabs, tabularx} 
\usepackage{setspace} 
\usepackage{multirow} 
\usepackage{rotating}
\usepackage[utf8]{inputenc} \usepackage[english]{babel} 

\begin{document}
\maketitle

\textbf{This paper accepted to the Journal ``Annals of the Brazilian Academy of Sciences" as part of the Proceedings for the BRICS Astronomy Workshop $-$ BAWG 2019 $-$, held in Rio de Janeiro, 29 Sep - 2 Oct 2019.}

\begin{abstract}
This paper is based on a proposal submitted for a BRICS astronomy flagship program, which was presented at the 2019 meeting of the BRICS Astronomy Working Group, held in Rio de Janeiro from 29 September to 2 October 2019. The future prospects for the detection and study of transient phenomena in the Universe heralds a new era in time domain astronomy. The case is presented for a dedicated BRICS-wide flagship program to develop a network of ground-based optical telescopes  for an all-sky survey to detect short lived optical transients and to allow follow-up of multi-wavelength and multi-messenger transient objects. This will leverage existing and planned new facilities within the BRICS countries and will also draw on the opportunities presented by other multi-wavelength space- and ground-based facilities that exist within the BRICS group. The proposed optical network would initially perform followup observations on new transients using existing telescopes. This would later expand to include a new global network of $\sim$70 wide-field 1-m telescopes which will cover the entire sky, \textit{simultaneously}, with a cadence of less than a few hours. This realization would represent a ground-breaking and unique global capability, presenting many scientific opportunities and associated spin-off benefits to all BRICS countries.
\end{abstract}

\section{Introduction}
Astrophysical transients are sources that suddenly appear for the first time, or known objects which brighten considerably, and then fade away, sometimes to oblivion. As we shall explain below, some transient events in the distant universe carry unique messages about fundamental laws of physics, while others closer to us have a much more direct impact on safety in our daily lives.


\begin{figure}[t]
    \centering
        \includegraphics[width=1\textwidth]{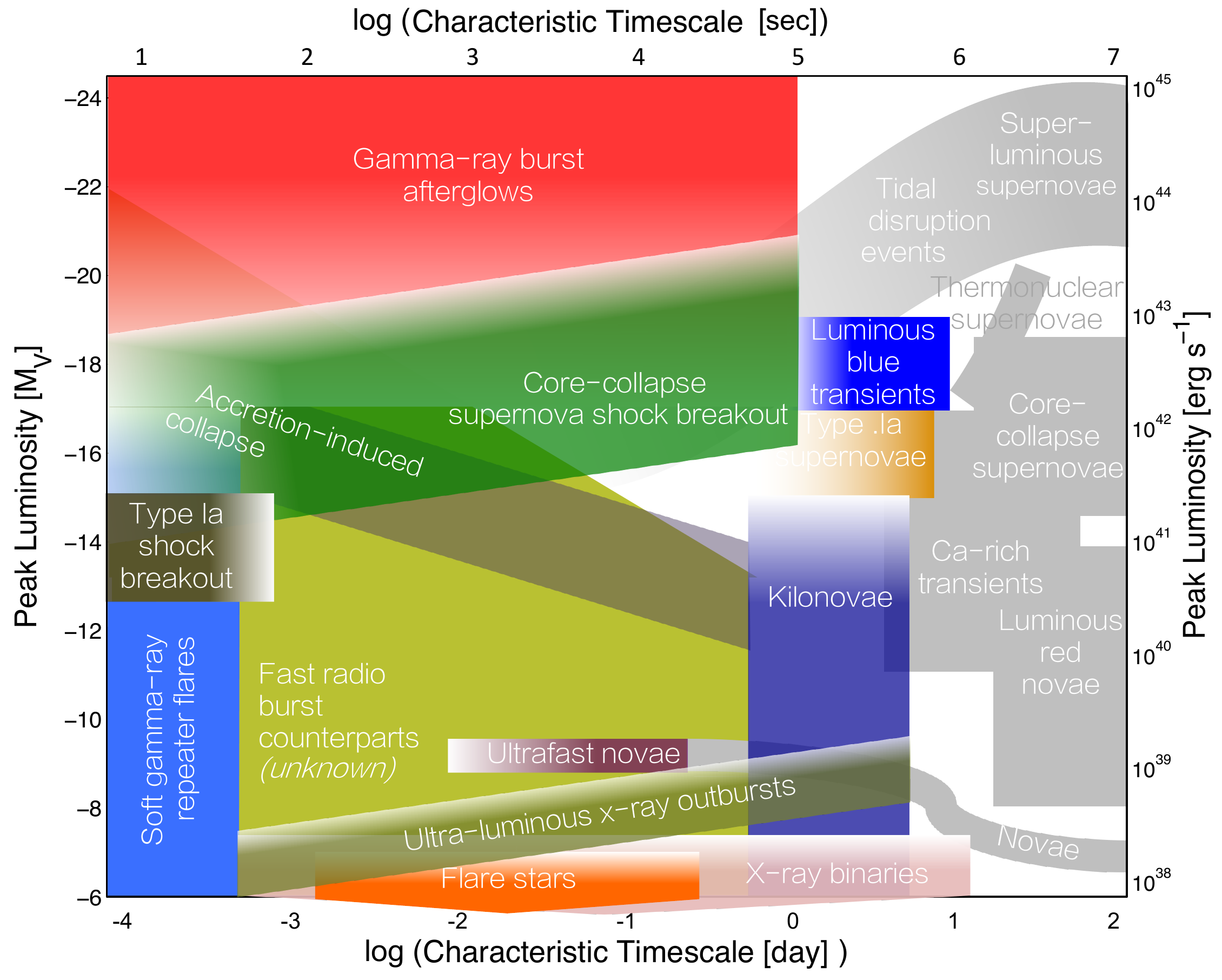} 
        \caption{Luminosity$-$Timescale plot for transient phenomena, with approximate characteristic timescales and peak luminosities (credit: Jeff Cooke/Swinburne University of Technology)}
        \label{fig:1}
\end{figure}

Transient phenomena are seen across a wide range of object classes, from the nearest to the farthest, and covering a wide range of luminosity, some being the most energetic events in the Universe. The latter often hold the key to our understanding of universal physical laws in conditions of extreme energy. We know of their existence because we have detected powerful bursts of gravitational waves, gamma rays, X-rays, or radio waves; however, we still know very little about their appearance in the optical bands. In addition, optical observations of transients may lead to important discoveries in other fields of astronomy: nearby asteroids potentially dangerous for Earth, Near-Earth Objects (NEOs), new extra solar planets; variability of super massive black holes in galaxies; stellar flares; gravitational lenses; and probably several other yet-unknown types of fast-changing phenomena that have not yet been discovered. Time domain astronomy, where repeated observations can be undertaken over a range of timescales (from sub-second to seconds, minutes, hours, days, months, years), is key to obtaining data of sufficient cadence to unlock the nature of transient objects. We show the ``discovery space'' for transients in Figure \ref{fig:1}, in terms of luminosity and typical timescales of variability.

This first phase of this program aims to leverage existing and future facilities within BRICS to develop an intelligent network of telescopes which can automatically conduct follow-up observations of newly discovered astrophysical transients. It will harness software systems involve event broker and to select different types of targets, then schedule and conduct observations. The second phase will be the development of a new BRICS-wide network of wide-field 1-m telescopes that, collectively, will cover all 4$\pi$ steradians of sky to a depth of r' $\sim$20 -- 21 and with a cadence of $<$ 1 hour.

\section{Scientific Rationale}
Despite all the technological developments and improvements in telescopes, instruments and detectors over the last few decades, we still know very little about transient events in the optical band on timescales of less than a day or so (see Figure \ref{fig:1}).  Theoretical models predict the appearance and disappearance of various classes of celestial transient sources on timescales of seconds to hours. We have already discovered signals from such short-lived events with X-ray, $\gamma$-ray, and radio telescopes, and more recently with multi-messenger gravitational wave and neutrino detectors. However, obtaining simultaneous optical information remains challenging, due in part to a lack of optical telescopes distributed around the world that can monitor the whole sky with a sufficient cadence (many times per night) and at sufficient depth to catch the optical counterparts of such transient events. A few transient observation programs are already underway in some BRICS countries and these will gradually expand with new facilities and systems over the next few years. Examples include the SALT transient program (begun in 2016; \cite{Buckley2015}); the ThunderKAT radio transient program (began in 2018; http://www.thunderkat.uct.ac.za/), employing the first dedicated optical telescope (MeerLICHT; http://www.meerlicht.uct.ac.za/) aimed to look for optical counterparts to radio transients; the MASTER Global Robotic Transient Network (\cite{Lipunov2010}), the EWACS program in China and the GROWTH program (http://growth.caltech.edu), with facilities in India and China. 
\begin{figure}[t]
        \centering
        \includegraphics[width=0.8\textwidth]{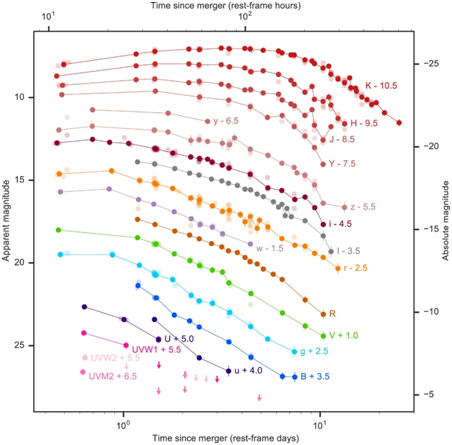} 
        \caption{Light curve evolution of the kilonova associated with GW170817 / GRB 180817A \cite{2018ApJ...855L..23A}}
\label{fig:2}
\end{figure}

In order to advance our understanding of transients we need a network of globally distributed wide-field telescopes that can rapidly scan the whole sky (many times per night) and with sufficient depth to catch as many transient events as possible. Other future programs will survey the sky and look for transients, but they will not have all the capability we are aiming for. For example, the forthcoming 8.4-m Rubin Observatory's Legacy Survey of Space and Time  (LSST) will efficiently survey much of the southern sky to a great depth, from 2023 on. However, its survey cadence will be of a few days, thus missing or poorly sampling many classes of fast transients. Moreover, the Rubin Observatory LSST consists of only one telescope, located in Chile; it will miss all the fast transients that happen when it is daytime in that single observing location. Finally, the LSST will cover the whole southern sky but almost none of the northern sky.    

Our proposed BRICS Flagship Program on astrophysical transients aims to address the lack of complete sky coverage at sufficiently high cadence by developing a distributed optical transient detection and followup network. This network will use existing and future facilities within the BRICS countries, or for which the BRICS countries have access. Such a network will allow for both the independent discovery and rapid followup of newly discovered transients from many sources, including the network itself, the Rubin Observatory LSST and other multi-wavelength and multi-messenger alert facilities, both ground-based and space-based.

\subsection{Synergies with transient searches at other bands (UV, X-ray, $\gamma$-ray)}
Our BRICS Optical Transient Network (BRICS-OTN) will integrate and complement transient searches in other electromagnetic bands, in particular at higher energies, allowing for contemporaneous and continuous uninterrupted optical observations, at a high cadence.

Three of the BRICS countries have their own astronomical satellites which allows for the detection of ultraviolet and X-ray transients (e.g. \textit{AstroSat}: India, \textit{Insight (HXMT)}: China, \textit{Spectrum-RG}: Russia). Additionally, Russia has a preliminary 25\% quota of the observing time for the \textit{INTEGRAL} observatory, which operates in hard X-rays and gamma rays. More are also in the process of development, for example \textit{WSO} in the UV  and  \textit{Millimetron} (Russia) in sub millimeter wavelength, \textit{Einstein Probe (China)} in X-rays. 

Time-domain astronomy is relatively unexplored in the UV, although it can have a high potential scientific impact on the study of transient events \cite{2014AJ....147...79S}. The last decade has been witness to the importance of the UV space missions in detecting transients in UV, with serendipitous \textit{GALEX} discoveries of transients.  However, neither \textit{GALEX} nor the currently operating UVIT instrument on \textit{AstroSat} were intended for observations of transients. Future UV missions within BRICS (\textit{WSO, SVOM} and smaller payloads planned by India (\cite{2018SPIE10699E..3EA,2018ExA....45..201M}) will address the lack of UV transient detection capabilities.

For X-ray transients, the most important synergy of the BRICS Optical Transient Network will be with the newly launched (July 2019) Russian-German \textit{Spectrum-RG} satellite and the planned Chinese-French \textit{SVOM} satellite (launch early-2020s). 
The \textit{Spectrum-RG} payload includes two telescopes: eROSITA (Germany) and ART-XC (Russia). The main scientific goal of the latter one is to map the Universe in the hard X-ray band (6--30 keV) with unprecedented sensitivity. It will lead to the detection of thousands of obscured AGNs and reveal many new Galactic sources, such as high-mass X-ray binaries and cataclysmic variables. It will also conduct several wide-field surveys of the most interesting regions of the X-ray sky, such as the Galactic Center, Norma Arm, Small Magellanic Cloud, over a 5-30 keV range. \textit{SVOM} will detect GRBs with its X-ray/$\gamma$-ray and optical instruments.
Finally, some of the BRICS countries also have (or will have) access to ground-based Cherenkov TeV $\gamma$-ray telescopes, like \textit{H.E.S.S.} and the \textit{Cherenkov Telescope Array (CTA)}.

\subsection{Major Science Drivers}
What are the major science drivers that we hope to address with BRICS-OTN? We split them into two broad groups: 1.) events that can tell us something more about fundamental physical and astrophysical processes; and 2.) events that can have a more immediate impact on humanity, namely its future existence. \\

\noindent
\textbf{1.	 Fundamental knowledge of physical laws and astrophysical processes in the Universe} \\
Transient events in this class are created by the sudden release of a huge amount of energy from a small region of space: those flashes briefly light up the universe, and can be visible from billions of light years away. For example, the merger of two neutron stars releases in a few seconds, and from a region of only about 50 km across, the same amount of energy radiated by the Sun over its whole lifetime (10 billion years). These events are the best experimental way for us to probe regions of extremely strong gravitational fields, electromagnetic fields, and matter density. They give us a chance to test the predictions of general relativity and nuclear physics, and witness the formation of black holes. 

In many cases, transients initially emit highly energetic particles and $\gamma$-rays, some of which get converted to optical photons in the surrounding plasma. Only by comparing the brightness evolution in different bands during the first crucial minutes and hours can we understand and model the physical processes that caused a transient burst. Moreover, optical observations usually provide a much more precise localization of the event than gravitational wave or gamma ray detectors; without an optical detection, it would be very hard to determine which galaxy is the host of a transient event.  The main types of high-energy transients that we aim to detect include the following:\\
\\
\begin{figure}
        \centering
        \includegraphics[width=1.0\textwidth]{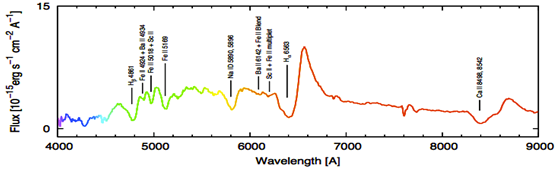} 
        \caption{Optical spectrum of a Type II-P supernova, SN 2017gmr, obtained using AD-FOSC on the ARIES 3.6-m DOT}
        \label{fig:3}
\end{figure}

\textbf{\textit{i)		Neutron star - neutron star mergers:}}\\
Neutron stars are the end result of the evolution and collapse of massive stars. When two neutron stars orbit each other in a tight binary, they tend to lose energy and spiral in via the emission of gravitational waves, until they finally collide and merge. The LIGO gravitational wave observatory (see \cite{Abbott2016} for details) has revealed the last stages of a binary neutron star merger in August 2017 with the  GW170817 event \cite{Abbott2017a}  Optical telescopes around the world scrambled to look for the visible flash associated with the merger (see Figure \ref{fig:2}) and featured in a large multi-wavelength study  (\cite{Abbott2017b}). It was eventually found after 11 hours. With the automated monitoring system we are proposing, we will detect and follow future optical counterparts of neutron star mergers during the first few hours, greatly enhancing our understanding of the event. For example, measuring the optical rise and peak in the first few hours is crucial to distinguish between radioactive decay of neutron-rich material and thermal cooling of the shocked ejecta (\cite{2017Natur.551..210A, 2010MNRAS.406.2650M})\\
\\

\begin{figure}
    \centering
        \includegraphics[width=0.8\textwidth]{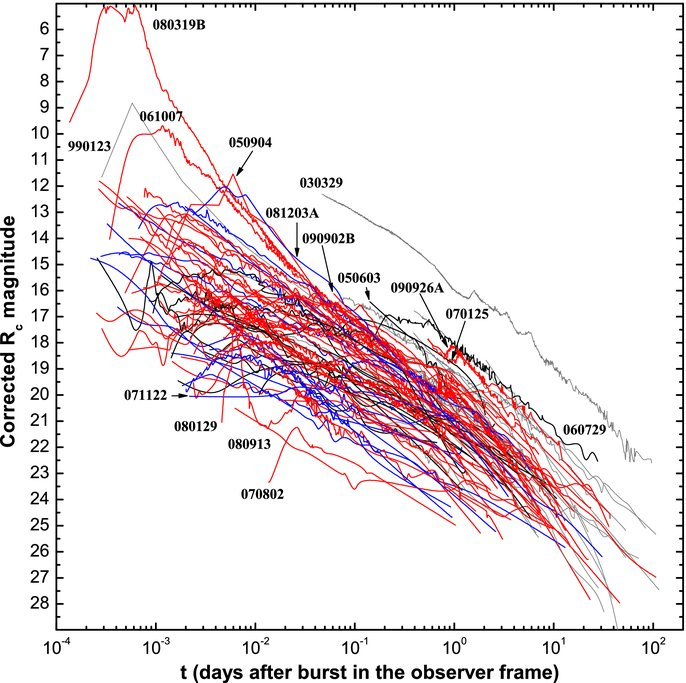}
        \caption{Optical afterglows of long  GRBs. \cite{2010ApJ...720.1513K}}
        \label{fig:4}
\end{figure}

\textbf{\textit{ii)		Core-Collapse and Super-Luminous Supernovae, Long $\gamma$-ray Bursts \& Fast Transients:}}\\
Stars more massive than about 8 times the mass of the Sun end their lives with an implosion of their iron core and explosive ejection of the envelope: a supernova (e.g. \cite{2010ApJ...725..904N}). A spectrum of a core collapse supernova is shown in Figure \ref{fig:3}, obtained with the newly completed 3.6-m DOT in India. Several scenarios are still the subject of theoretical debate. For example: what fraction of supernovae also produce a collimated jet, seen as a Long Gamma Ray Bursts (GRBs); and whether there is a maximum mass above which stars directly collapse into a black hole without a supernova explosion. Transient networks can also assist in providing long-term light curves, for example in Figure \ref{fig:4} we show the ensemble of long GRBs light curves obtained from a variety of telescopes networks.

The enormous developments in transient research in the last decade has led to the discovery of new kinds of rare events, which are 10--100 times more luminous than canonical core-collapse supernovae (CCSNe). These are Super-luminous Supernovae (SLSNe; e.g. \cite{Dong2016}) which spectroscopically can be categorized as hydrogen-rich (SLSN-II) and hydrogen deficient (SLSN-I) events. The powering mechanism of SLSNe is not yet resolved. Interaction of SN-shock with circumstellar medium (CSM), the presence of a spin-down magnetar or pair-instability supernovae (PISNe) are the current proposed theories. The progenitors of SLSN-I events may have a connection with energetic stripped-envelope CCSNe and Long-Gamma Ray Bursts (GRBs). In fact, transients with intermediate luminosity (between CCSNe and SLSNe) have also been found. Another proposition is SLSNe and CCSNe do not belong to the same population. Thus as a whole, the origin of SLSNe is yet unknown. 

In the last 4$-$5 years, we have also found a few energetics transients ($<$10) which rise very fast and can attain a peak luminosities comparable to the energetic of CCSNe or SLSNe. These are called fast-transients (FT) and their origins are also unknown. The light curve evolution of these events can be modeled with spin-down magnetar scenario or a merger-system of Neutron Star-White Dwarf, or accretion induced collapse of White Dwarf model. The spectral evolution of such events vary significantly from case to case. They may also have a close connection with kilonovae, which has been found to be a source of short-GRBs and associated gravitational wave (GW) events (i.e. for GW170817).\\
\\

\textbf{\textit{iii)	Optical and ultraviolet flares in galactic nuclei:}}\\
Recent research in the field of time-domain astronomy (e.g., the iPTF/ZTF, ATLAS, ASAS-SN surveys) have revealed several new types of transients near the centers of galaxies. Most of them are brighter than canonical supernovae (SNe) and exhibit very broad light curves with almost featureless spectra at early stages. They are either tidal disruptions (TDEs; see \cite{Holoien2020} and reference therein) of solar-mass stars by nuclear super-massive black holes, or luminous supernovae (SNe) in the star-forming nuclear region. Some of the transients are optically bright but not detected in X-rays, while others are X-ray bright and optically faint.

TDEs are rare transients which occur when a star is tidally disrupted by the central black hole in a galaxy. Most galaxies and perhaps some globular clusters contain a massive black hole at their centre. Such black holes are usually inactive or very faint, for lack of accretion. However, if a star passes too close to a nuclear black hole, it gets shredded by tidal forces, and its gas forms a transient accretion disk around the compact object. For a short time (over $\sim$10's of days), the black hole becomes active again, giving us a chance to study its properties.
Numerical simulations have shown that during a TDE event about half of the star's mass is unbounded and the remaining half forms an accretion disk.  Because of the thermal emission produced by the accretion disk a bright transient is observed that can last from a few days to a few weeks to even months before fading. Future TDE discoveries from UV or X-ray missions within BRICS could be effectively followed up by the BRICS-OTN. 
High cadence optical light curves of TDEs will probe the size of the accreting region and the mass of the black hole. More systematic optical and multi-band monitoring of nuclear transients, as afforded by BRICS-OTN, would greatly advance our understanding of the extreme physics happening at the centers of the galaxies.
With the frequent all-sky monitoring, we expect to discover many such transients every year. We will also probe to much further distances than is possible with the typically smaller aperture all-sky-monitoring telescopes available today.\\
\\
\textbf{\textit{iv)	Optical counterpart of Fast Radio Bursts:}}\\
This mysterious class of transients has baffled radio astronomers for more than a decade. Dozens of radio bursts have been detected so far, with duration's of a few milliseconds, from unidentified extragalactic sources. Current wide-field optical monitors lack the cadence and depth to identify transient optical afterglows at the position of the bursts. Together with the radio telescope facilities at our disposal (e.g. MeerKAT, FAST, BSA) and those in the future (SKA),  our proposed BRICS Optical Transient Network will have the best chance to detect optical counterparts and to resolve, for the first time, the host galaxy and physical origin of this phenomenon.\\
\\
\textbf{\textit{v)		Reverberation mapping: estimating black hole masses in AGNs:}}\\
Reverberation mapping (RM) is a technique for studying the structure and kinematics of the broad-line regions (BLRs) of active galactic nuclei (AGN), including their most luminous subset, the quasars. RM is a particularly important tool as the BLRs generally project to angular sizes of only tens of micro-arcseconds or less, too small to be resolved directly by any current or near-future technology. RM experiments are arduous, as they require a large amount of telescope time and coordinated effort. RM measurements have been performed for about 50 AGN, most often only for the H$\beta$ emission line and almost exclusively for AGN at z $<$ 0.3. These studies have been used to derive the RBLR-LAGN relation for AGN, in order to calibrate the single-epoch super massive black hole (SMBH) mass estimation method, used extensively for determining the SMBH mass and the corresponding Eddington ratio of the AGN. However, it should be appreciated that the presently available AGN sample, for which reliable RM measurements have been made, are strongly biased towards very bright (i.e. L $>$ 10$^{43}$ erg/s) and highly variable AGNs, due to which these existing RM studies do not probe the AGN parameter space with any degree of uniformity. To remedy this situation, it is imperative to carry out the time-consuming RM programs on a network of moderate-size optical telescopes across the globe, like the proposed BRICS Optical Transient Network. 

The low luminosity AGN are likely to have low BH mass and hence their reverberation period will be smaller on time-scales of less than one day up to a few days. We would be targeting low luminosity/low mass AGN for such a program to fill the gap where low black hole masses have been seldom inferred by this method. The available resources comprising many of the existing telescope within the BRICS countries, and the eventual continuous monitoring potential of an all-sky network would enable telescopes at different longitudes to get a near-continuous observations of AGN. \\
\\
\textbf{\textit{vi)	Outbursts and state transitions in X-ray binaries}}\\
Neutron star and black hole X-ray binaries consist of a donor star transferring mass onto the compact object. The accretion flow is not steady: long periods of quiescence may be punctuated by luminous outbursts which last for several weeks or months. Although such outbursts are usually initially discovered in the X-ray band, simultaneous multi-band observations during the outburst are necessary to model the changing structure of the accretion flow and the nature of the accreting object. In particular, optical observations are particularly important to determine or constrain the type of donor star, the outer radius of the accretion disk, the presence (and subsequent disappearance) of synchrotron emission from a relativistic jet. Day-to-day and hour-to-hour variability in the optical emission can be caused by reprocessing of the variable X-ray emission in the irradiated outer disk, or by changes in the jet properties associated with sudden ejections. Our BRICS Optical Transient Network will enable continuous coverage of the optical light curve for the most interesting X-ray binary outbursts, particularly over the first few weeks when those sources typically switch from a jet-dominated to a disk-dominated accretion state. We estimate that we will follow (on average) a couple of Galactic black hole X-ray binary outbursts per year, and about a dozen neutron star outbursts. Our team has a long, recognized experience of multi-band photometric and spectroscopic studies in this field.\\

In addition to all of the above classes of transients, rapid all-sky optical monitoring may lead to important discoveries in other fields of astronomy: accreting binary stars (novae, cataclysmic variables, symbiotic nova, intermediate luminosity and red novae, etc.), stellar flares; gravitational microlensing; extra-solar planets (from transits or gravitational microlensing \cite{2012Natur.481..167C}); planetary collisions in other stellar systems detected in the UV \cite{2008ApJ...679L..41W}. But perhaps the most exciting possibility of the BRICS optical transient network is the serendipitous discovery of as yet new classes of transients, representing new phenomena and physical processes which, due to the short-comings of existing surveys (lack of complete sky coverage at sufficiently high cadence) have remained undiscovered.\\
\\

\paragraph{2: Practical benefits: detecting hazardous near-Earth objects}
In some cases optical transients are much more directly linked to humanity: they may come from potentially dangerous near-Earth objects (NEOs). NEOs are asteroids and comets that have been nudged by gravitational perturbations into orbits that penetrate the Earth’s neighbourhood with the possibility of an eventual collision. This is an obvious hazard problem. To address this issue, the first step is a reliable and advanced detection of approaching NEOs, followed by a characterization of their properties (trajectory, size, mass), a risk assessment, and if necessary, urgent measures for protection and mitigation of the impact consequences. Early detection is the most crucial link in this chain. 

The 20-m meteoroid that exploded over Chelyabinsk (Russia) on 15 February 2013 at about 9:20 a.m. was undetected by any space-based or ground-based observatories. More than 1600 people were injured and the total damage was $\$$25M. However, the potential threat of NEOs involves a far higher danger to humanity, with potential destruction of cities and the death of millions of their inhabitants. A similar event in 1908, which occurred in the remote underpopulated eastern Siberian taiga region of Tunguska, has recently been estimated to have an explosive power equivalent to 20 Megatons of TNT, enough to destroy a city.

One of the major lessons of the Chelyabinsk event is obvious: the need to detect 10 meter class NEOs with a network of wide field telescope to continuously monitor the entire sky every few hours. The proposed BRICS optical transient network will ensure detection of many more dangerous celestial bodies than can be done presently and with sufficient accuracy to ensure an early warning, not less than 4 hours, which is a minimum time required for civil defense mitigation.

While the vast majority (about 90$\%$) of all large ($>$1 km) potentially hazardous objects (PHOs) have already been identified and characterized by the USA Space Guard Program, threats remain for smaller objects (10's of meters).  One of the science drivers for the LSST is to address a USA Congressional Act, passed in 2005, to detect, track, catalogue, and characterize the physical characteristics of PHOs equal to or larger than 140 metres in diameter that come within 1.3 astronomical units (AU) or less from the Sun. 

In the next decade, the majority of NEOs with diameters between about 100 m and 1 km are expected to be identified by large and moderate aperture survey telescopes such as the Zwicky Transient Facility (ZTF), the Catalina Sky Survey (CSS), Asteroid Terrestrial-impact Last Alert System (ATLAS), the Panoramic Survey Telescope \& Rapid Response System (Pan-STARRS), Fly-Eye NEO Survey Telescope (from $\sim$ 2020),  and eventually the Rubin Observatory LSST (from $\sim$2023). 

However, these survey telescopes don’t have a high enough survey rate to detect all PHOs, particularly smaller bodies with sizes between 10 m and 100 m (\cite{Mainzer2011, Mainzer2014}). The number of potentially hazardous bodies of decameter sizes is estimated to be 10$^8$, with a rate of collision with Earth of about one event per 5–10 years. Other small-aperture systems (e.g. Evryscope, SuperWASP, MASTER, GWAC) can observe the whole sky very quickly (up to few tens of minutes), but they are unable to provide sensitivity sufficient to detect 10-m class objects at distances greater than a $\sim$ 1 million kilometres. 
The estimate of the frequency at which Earth gets impacted by small asteroids ranges between 1 every $\sim$200 years for 20-m sized object to 1 every couple of thousand years for a 50-m sized object (\cite{Stokes2019}). Also, the likelihood of the next Earth impactor being a 50m object is roughly 1000 times greater than it being $\sim$1 km in diameter because of the power-law size-distribution. Knowing the physical properties of the small-asteroid population is therefore vital since a 20-50m object can still cause town-sized to even metropolis-sized destruction depending on its composition, density, structural integrity, and impact angle.

The proposed worldwide BRICS network would meet the goal for successful detection and orbit determination  of NEOs down to 10's of meters in size over the full sky every few hours, reaching a depth of 20th magnitude. Our proposed global network of 1-m wide field (10's of square degrees) optical telescopes are ideally suited to do that purpose.  Our survey observation strategies will allow for dedicated follow-up of PHO alerts, allowing us to track the path of the incoming object and compute the most likely location of impact.\\

\begin{figure}[tbh] 
\begin{center}
    \includegraphics[width=1.0\textwidth]{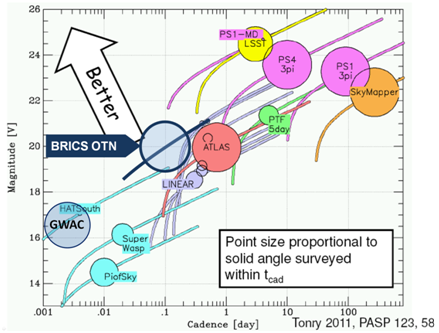}
	\caption {BRICS Optical Transient Network (BRICS OTN) survey merit diagram. Adapted from \cite{Tonry2011}.}
	\label{Fig5}
\end{center}
\label{fig:5}
\end{figure}

\section{Rationale for a BRICS Flagship program}
A BRICS Optical Transient Network, covering the entire sky continuously with a cadence of $\sim$ hours, will greatly expand the parameter space in which transients can be detected and studied (see Fig. 5). Using our proposed network of 1-m class telescopes, we will discover many transients that otherwise would be missed by more sky-limited or time-limited surveys. Such a network will result both in the discovery of new fainter transients over the $\textit{whole}$ sky, and in the more rapid and efficient detection of optical counterparts to transients originally discovered in other bands. We will monitor them with sufficient cadence to study their fast evolution in far more detail than has been possible in the past with individual smaller aperture facilities, many of which are not optimally connected for efficient and automated observations. The BRICS-OTN will also bring together, in a synergistic manner, all of the existing telescopes across the electromagnetic spectrum which are available to us within BRICS, plus international facilities for which we have access, allowing for coordinated multi-wavelength follow-up observations. It is from combined multi-band and multi-messenger studies that the deeper physical insights into the nature of transient phenomena are usually derived; moreover, such combined studies will make the best use of the diverse expertise and facilities in different BRICS countries and institutions.

Our program aims to enhance levels of collaboration within the BRICS group on multi-wavelength astronomy, focusing on astrophysical transients and time domain studies, including high-time-resolution monitoring. Our research will cover many classes of transient systems: some powered by accretion of gas onto compact objects and associated outflows (e.g., in relativistic jets); others caused by stellar collapses, mergers, or collisions. In all cases, we will be exploring the physics behind some of the most energetic events in the Universe. The program will build on the existing transient observation programs already underway in some BRICS countries and will expand to take advantage of newly created facilities (e.g., the FAST radio telescope in China, the MeerKAT radio array in South Africa and the Spectrum-R\"{o}entgen-Gamma satellite launched by Russia), and of future facilities (e.g., the SKA at radio bands, the LSST in the optical band, the CTA for gamma rays). In addition, BRICS-OTN will be a ground-breaking tool to search for electromagnetic counterparts of multi-messenger transient detected with LIGO/Virgo (gravitational waves) and Ice-CUBE (neutrinos). 

The collaborators named in this proposal (and many others not named, who will have plenty of opportunities to join later) have a wide expertise and direct experience at some of the world’s forefront facilities, including those within BRICS: the future success of our project rests on the proven quality of our team. Coordinating a BRICS-wide program on transient astronomy, bringing together people and machines, could really have a major global influence on this fast growing field. The main stepping stone towards the successful implementation of a worldwide network will be the experience gained from LSST, for which 4 of the BRICS countries are already involved as international collaborators. LSST is due to begin routine survey operations in 2023, and its data production will be enormous, with an estimated several million transient detections per night. The manageable data stream astronomers are currently used to (e.g., from individual telescopes with a small field of view) will turn into a veritable fire hose when LSST comes of age, and this will necessitate a new approach for the management of such ``Big Data''. This aspect is covered by a parallel BRICS Flagship Program on Big Data challenges (PI: R. Taylor) which will address some of the needs for the management of LSST data and will develop new data-analysis solutions within BRICS. The techniques we currently use in the study of transients today will be wholly inappropriate with such enormous data volumes, which only Big Data and Machine Learning techniques will be able to cope with.  But the training datasets required for efficient data processing in these facilities will have to come from our current experience of transient discoveries and prompt follow-up studies. Managing LSST transient data is a challenge that epitomizes the ``4th Industrial Revolution" (4IR), and both the BRICSKA and BRICS-OTN programs will fully exploit Artificial Intelligence, Deep Learning and the many developing techniques applicable to the challenges of the 4IR.

A BRICS-OTN will enhance international scientific cooperation, particularly within the BRICS group, and promote interest in physics, information technology and engineering, with positive spin-offs outside astronomy, including human capacity development and outreach. The network will enhance levels of collaboration within the BRICS member countries in multi-wavelength astronomy and the emerging field of multi-messenger astronomy (e.g., detection of gravitational waves and neutrinos), both underpinned by time domain astronomy capabilities. Collaborative intellectual investment will be required in order to develop the most efficient connections between human capabilities, observing facilities, and output data streams.

The BRICS member country’s participation as global leaders in astrophysical transient research is expected to accelerate greatly over the next decade with the development of new facilities. A cornerstone of this will be the emergence of the LSST as a generator of transient discoveries. Two of the BRICS countries (Brazil and South Africa) are now actively participating in LSST science collaborations and more may still do so in the future. Key to success is the bringing together the many impressive multi-wavelength followup facilities within BRICS. Examples of specific facilities for which many BRICS countries are already, or will be, involved are: MeerKAT, FAST, GMRT, SKA, BSA (radio), PRIME, IRSF (infrared), MeerLICHT, MASTER, ATLAS II, all the BRICS national observatories (optical), WSO (UV), AstroSat (UV \& X-ray), HXMT/Insight, INTEGRAL, Spectr-RG, SVOM (X-ray) and HESS, CTA, SVOM ($\gamma$-ray).

\section{ Objectives \& Methodology}
Our primary aim is to bring together scientists and engineers within the BRICS group to collaborate on one of the most compelling topics in astronomy and astrophysics:  time domain studies of astrophysical transients and variable objects. We aim to harness the existing optical facilities within the BRICS countries in a cohesive network of telescopes which can efficiently detect and perform follow-up observations of transient objects of many classes. The cross-cutting nature of the program will allow synergies to evolve throughout the BRICS family, which will drive forward both the pure science knowledge and, equally important, the development of human capacity in science and technology, through the training of students. In addition, the program will build on the multi-wavelength and multi-messenger facilities that exist within BRICS, or for which BRICS scientists have access. Transients discovered by such facilities will then be more easily studied at optical wavelengths with the proposed BRICS Optical Transient Network. 

\subsection{Use existing facilities}
The program will start by networking our existing optical facilities within BRICS to allow for automated follow-up of transients. Such facilities include an array of about 12 $\times$ 1-m-class (1.0--1.5 m) telescopes, 5 $\times$ 2-m-class telescope (1.6--2.4 m), one 3.6-m telescope (DOT; India), access to the 4-m SOAR (via Brazil), to the 8-m Gemini telescope (via Brazil) and the 10-m SALT (South Africa). The 1-m class telescopes will primarily be used for photometric monitoring, or intensive high-speed photometry, and will be networked together for continuous all-sky coverage. The larger telescopes will primarily support spectroscopic observations of particularly compelling transients, selected as target-of-opportunities from the photometric monitoring.

Event brokers coupled with Target and Observation Tools (TOMs) will be employed to select and observe compelling targets, based on their scientific importance and on the availability and technical capability of follow-up facilities. We are cognizant of the fact that most BRICS facilities service a wide range of users, who are conducting many different programs. The idea of the BRICS network using the existing facilities is to provide $\textit{opportunities}$ to conduct followup observations in an automated and efficient manner, \textit{should it be considered that such observations are of sufficient scientific value.} These activities would begin immediately and take several years to complete.

The SAAO's \textit{African Intelligent Observatory (AIO)} in South Africa (PI: S. Potter) is a new project to network \textit{existing} telescopes to enable the automated follow-up of transients.  As described below, the project has the potential to become a BRICS-wide initiative, involving many facilities, which would present a major engineering and software project, building on and expanding on capabilities in technology development within all the BRICS countries.

The SAAO telescopes in Sutherland are the premier optical astronomical facilities in Africa.  They are available to the entire South African astronomy community, as well as to international researchers, including from BRICS.  Several other Sutherland telescopes operate under collaborative programs with international partners and function in a variety of modes from automated all-sky surveys to queue-scheduled observations (e.g. SALT). This diversity ranges from wide-angle imaging, high-speed photometry and spectroscopy to niche specialisations such as spectropolarimetry and Fabry-Perot imaging.

The primary science driver for the African Intelligent Observatory (AIO) is time-domain and transient science. Critical to the success of transient science is the rapid and intelligent use of Sutherland telescopes to react to triggers/alerts from other ground- and spaced-based observatories. Triggers from LIGO/Virgo and MeerKAT will likely be the highest priority initially for the SIO, with SKA, CTA and LSST providing opportunities in subsequent years. Also, at the same time, the project will substantially increase the efficiency of all observing programs through automated observations of other types of targets. The vision for the Sutherland Observatory is to eventually have all SAAO telescopes tied into the SIO network. Multiple hosted facilities are also anticipated to contribute, providing triggers and/or follow-up observations. 

To fully capitalize on time-domain and transient science, intelligent algorithms are needed to filter the deluge of transient candidates in the primary survey, and select only the most interesting and relevant targets to follow up.  Determination of the magnitude of a target is best done with optical imaging.  Identification and characterization of unknown-origin triggers are best done with a relatively low-resolution spectrograph, targeting point sources with a single exposure, with the highest possible throughput, and a broad optical/IR wavelength coverage.

Under the AIO, the various optical/IR facilities at Sutherland will be tied together into a network that can efficiently carry out standard observations and respond to triggers autonomously.   The system will work collectively and intelligently, satisfying the needs both of standard observing and of fast, efficient responses to triggers.  The full ambition of the project includes utilizing all telescopes and instruments, including SALT. Some of these instruments are still under development and need to be completed. Some telescopes require hardware modifications in order to be remotely, and then robotically, operable as well as to allow instrument selection. Software must be developed to run the processes and control all of the hardware. The SIO project is in line with the South African National Multi-wavelength Strategy (2015), the recommendations of the SALT External Review Committee (2016), and the South African Strategy for Optical Astronomy and Next Generation Instrumentation for SALT (2017). In addition, the BRICS-OTN program may gain access to $\sim$10\% of SALT time for followup observations through purchase of telescope time each semester (or possibly joining as a SALT partner). 

\subsection{Develop New facilities}
Longer term (5$-$10 years) we aim to develop a \textit{dedicated} network of globally distributed 1-m class telescopes within all of the BRICS countries which would provide the enhanced capability of all-sky coverage at high cadence, as described in previous sections of this proposal. This future development builds on some existing initiatives within BRICS:  \textit{Sitian} in China (PI: J. Liu) and \textit{PHOBOS} in Russia (PI: B. Shustov).

$\textit{Sitian}$  is perhaps the most ambitious of these initiatives, namely a globally distributed  system of optical telescopes allowing for all-sky monitoring (northern and southern hemispheres). It is proposed to consist of many (24 groups of 3) 1-m wide field (5$^{\circ}$ $\times$ 5$^{\circ}$) Schmidt telescopes which, when combined, will image the entire sky every $\sim$30 min. Each field will be observed simultaneously in 3 filter  (\textit{g, r, i}) to a depth reaching 21.5 mag. The telescope will be controlled by a central ‘brain’,  which determine the observing schedule, react to fast alerts and follow-ups, control each observation, handle the data management and processing. Future developments of \textit{Sitian} could include provision for spectroscopic followup with dedicated 4-m class telescopes and and expansion of the network globally.

\subsection{Development goals}
The BRICS optical transient network program will involve activities to assist in the development goals within the BRICS countries, covering both human and technology development. We anticipate the following general activities to be conducted as part of the program:

\begin{itemize}
\item{workshops on transient data handling; data reduction tools; Virtual Observatory tools; accessing public multi-wavelength data archives}

\item{science workshops and meetings for a wider potential community of participants}

\item{programs for students, such as summer/winter schools and internships at BRICS host institutes}

\item{development of software engineering capacity within BRICS partners through the following collaborative activities:}

\begin{itemize}

\item{designing and building a series of Telescope and Observation Tools (TOMs) to control telescopes/instruments with a network}

\item{developing pipelines for searching and identifying optical transients}

\item{developing or modifying event brokers to select transient events for follow-up with BRICS-OTN.}

\end{itemize}

\item{capacity development in engineering applicable to design and construction of new telescopes within the BRICS-OTN (e.g. optics, detectors, telescope design, control software}

\end{itemize}

\section{Feasibility}
The BRICS OTN will begin by utilizing \textit{existing} facilities within BRICS, requiring investment primarily in new software development and associated hardware necessary for updating existing telescopes (e.g. motion control systems). Since this will involve facilities at existing observatories, the basic infrastructure and hardware already exists. This first phase will therefore primarily require manpower efforts, estimated at 5 FTEs distributed within the 5 BRICS countries. We expect that this software engineering team will be capable of undertaking all of the tasks to complete the networking of existing telescopes, in a phased approach, over a period of $\sim$3 years. The SAAO will be capable of providing some leadership or assistance, as needed, since it will already have embarked on the similar IO program. Furthermore, for the development of the Telescope and Observations Managers (TOMs), the software tools which can control and initiate observations, we expect to leverage the existing support available through a developing collaboration between SAAO and Las Cumbres Observatory (LCO) in the USA. 

The automated event driven transient followup systems envisaged for the BRICS-OTN will be designed to be scaleable, allowing additional telescopes to be added to the network seamlessly. In addition, from the outset we will design a survey option for controlling the telescopes to optimally cover the entire sky from both hemispheres. This will allow for BRICS-OTN to search for optical transients and also to patrol fields which are being observed by other multi-wavelength/multi-messenger facilities.

For the second phase of the BRICS-OTN program, namely the development of \textit{new facilities} to be distributed amongst existing observatories within BRICS, there are several key elements involved, which are outlined here:

The first key element of new facilities of the BRICS Optical Transient Network is the cost effective design of a networked system of wide field telescopes which can start with a small number and grow. The top-level requirements for the BRICS-OTN are as follows:
\begin{itemize}
    \item{The survey rate more than 5000 deg$^2$ hr$^{-1}$. This survey rate is required to observe the whole available sky every few hours.}
    \item{24h operation time. This requirement is important to ensure that the system will be able to respond to any alert and any given time on night sky.}
    \item{Limiting magnitude of at least $AB\sim21$}
    \item{Range of efficient science filters}
    \item{Fast readout cameras}
    \item{Angular resolution better that 2 arcsec is required (TBC)}
\end{itemize}

In 2017 INASAN built a new cost-effective 1-m telescope (AZ1000WF) with a 3$^{\circ}$ field of view and modern a focal unit design to operate with largest available CCD camera of 10.5 $\times$ 10.5 kpixel, together with a large filter set. This telescope could be considered as prototype for BRICS optical transient network. The parameters of the proposed new telescopes could  be improved in terms of the field of view (FoV) and detector, e.g.  the FoV can be increased up to 4 degree in case of use prime focus corrector and up to 7 degree in case of use Schmidt optical scheme. Other proposals of more efficient and wider field telescopes are considered by BRICS partners, particularly the designs proposed by NAOC for the \textit{Sitian} telescopes, currently based on a Schmidt design with a 25 square degree FoV. 

There already exists expertise in telescope design and construction within BRICS, including mirror fabrication (e.g. Russia, India and China). Some of this is leveraged through Indian and Chinese involvement in the Thirty Meter Telescope (TMT) project where both countries are contributing mirrors and in the case of India, the actuators and edge sensors.  China has full capability to fabricate the new telescopes for BRICS-OTN, for example through the Nanjing Institute for Astronomical Optics and Technology (NIAOT). All BRICS countries have the mechanical engineering design and fabrication capabilities for the design and construction of the telescope mounts.  

The second key element of optical transient network is efficient detectors. Russia has experience in building a modern camera based on new scientific CMOS devices (sCMOS) with a size of 61 $\times$ 61 mm, sourced from the Chinese company GPIXEL. This camera has advantages over existing classical CCD camera in terms of readout noise and readout speed, which is important for optical transient network. 

To operate with the new 1-m class telescope for BRICS optical transient network, the photosensitive area of the camera can be increased up to 92 $\times$ 92 mm using custom sCMOS from GPIXEL and up to 190 $\times$ 190mm for a mosaiced option.

The existing filters set based on the INASAN AZ1000WF telescope and camera could be improved using mosaic filters and we could also employ grisms for low resolution R=100 slitless spectroscopy. This option could be very important for spectrophotometry measurements and quick definition of the faint objects properties.

The third key element of optical transient network is the planning managing, processing and curating the data produced by the BRICS-OTN. Construction of a project data center (possibly a distributed platform) is one option to be considered. This may form part of the parallel BRICSKA program, which is planning to host both data from the SAAO's African Intelligent Observatory (AIO),  plus LSST data accessed by South African investigators and possibly for the wider BRICS community.

\section{Strategic Goals}
Here we discussion how the program aligns with the strategic goals of the respective BRICS countries:

\medskip\noindent
\textbf{\em Brazil}: \\
Brazilian astronomy has a large tradition in the study of variable astronomical objects. There are presently many initiatives, involving Brazilian collaborations, that can benefit the study of transient objects. Some of them are cited below. Various Brazilian institutions - e.g. Laboratório Interinstitucional de e-Astronomia (LIneA), Laborat{\'o}rio Nacional de Astrofísica (LNA), University of São Paulo (USP), Observatório Nacional (ON) - are putting a significant effort in various aspects of wide-field surveys like J-PAS, J-PLUS and LSST. In particular, the S-PLUS project is a wide-field survey completely led by Brazil. In addition, Brazil is also partner of the Cherenkov Telescope Array (CTA) project and has also access to large optical telescopes (SOAR and Gemini) as well as a range of 1-2m class telescopes. In particular, the ``Gemini In The Era of Multi-Messenger Astronomy" (GEMMA) program is highly focused on transient objects. Brazil has also expertise on high-energy studies of variable objects, mainly interacting binaries. There is a strong interest in Brazil in NEOs and solar system astronomy. All these projects and expertise provide a large potential of synergy inside BRICS communities. Brazil has a major contribution to the development of astronomy in South America, which could be boosted by a collaborative effort among BRICS countries.

\medskip\noindent
\textbf{\em Russia}: \\
One strategic goal is to incorporate the existing Russian infrastructure (through the creation of the proposed network) in the optical and $\gamma$-ray/X-ray bands in support of leading international projects such as LIGO/Virgo/KAGRA, LSST, SKA and SVOM. Using the available experience of creating our own wide-angle telescopes to answer the challenges of modern astronomy by observing the entire sky with high temporal resolution. Another goal is the creation, within the network, of software pipelines for searching, cross-matching and identifying new transient sources in the optical. This will be based on machine learning (ML) methods, using data obtained within the framework of the project. The tactical goal is to have an access for  astronomical observations in the southern sky.

The Spectrum-Roentgen-Gamma (Spectrum-RG) observatory was successively launched on 13 July 2019. In the framework of this project we propose to perform follow-up optical observations of newly discovered or known transient sources, registered with the ART-XC instrument on Spectrum-RG. Optical observations with BRICS-OTN could form part of a global program of the identification of sources, detected by ART-XC in its all-sky survey. 

\medskip\noindent
\textbf{\em India}: \\
The Indian government has made a substantial investment into astronomy with the {\em Thirty Meter Telescope, SKA, and LIGO} along with existing observatories in Devasthal and in Hanle. These observatories have made major contributions to the detection and monitoring of transient sources, particularly with new facilities such as the {\em GROWTH} facility. This may also eventually culminate in India's own 10-m class telescope (NLOT), which would have a major impact on transient followup as provided by the BRICS-OTN. Transients are an important science contributor to India's space observatories, both existing (AstroSat) and those in planning (DAKSHA). This proposal will be important in coordinating efforts by Indian astronomers with the broader BRICS community, with its world-wide coverage and range of available facilities. India is also a participant in the LSST project.

This program will require strengths in automated analysis of large amounts of data using the latest artificial intelligence techniques to mine through the data. This is of great importance to the country as India moves to new areas of innovation where there are already core strengths.

\medskip\noindent
\textbf{\em China}: \\
Global astronomy collaborations, as the one envisaged with our proposed BRICS-OTN,  aligns well with the theme ``building a community of common destiny for mankind", as proposed by the Chinese government. China has put large investment in technology research, culminating in many new astronomy facilities (e.g. LAMOST, FAST, HXMT (\textit{Insight}), and more to come (e.g. \textit{Xuntian}, the Chinese space telescope, \textit{Einstein Probe} and \textit{SVOM}). In parallel to this, optical astronomical telescope and instrument development in China has grown significantly, particularly with the development of the Nanjing Institute of  Astronomical Optics and Technology, NIAOT. This facility, and others in China, are expected to play a major role in the design and fabrication of new telescopes which will comprise the future expanded BRICS-OTN. Chinese participation in the \textit{Thirty Meter Telescope (TMT)}, will enable further leverage in technology development that would be beneficial to the BRICS-OTN. This will also eventually culminate in China's own 12-m telescope, which would also be a major role player in the followup of transients discovered by the BRICS-OTN.

China, through the Chinese Academy of Sciences, operates astronomical facilities which are already engaged in transient studies and for which some of these facilities could be inter-linked into the first phase of the BRICS-OTN. The National Astronomical Observatory of the Chinese Academy of Science (NAOC) operates a number of 1$-$2.4 optical telescopes as well as the FAST and other radio telescopes. In addition, through IHEP, China also has an existing X-ray satellite (\textit{Insight}), and more planned for the future (e.g. \textit{SVOM}). Such facilities will hugely benefit from the supporting BRICS-OTN, which could provide almost continuous monitoring of multi-wavelength transient events. China is also a potential participant in the LSST project, where transients is one area of interest.

\medskip\noindent
\textbf{\em South Africa}: \\
The program closely aligns with the National Strategy for Multi-wavelength Astronomy, building on the large investment South Africa has made in astronomy through SALT, MeerKAT and in the future with SKA and LSST. Exploiting the LSST alert stream will allow South Africa and other BRICS countries who have access to actively participate in transient follow-up. This program seeks to enhance international science collaboration by combining its own facilities with those within BRICS. Development of new instruments, software and human capacity within the field of transient astronomy is one of the goals of the project which will be actively supported by SAAO.  

This program aligns with the National Strategy for Human Capacity Development for Research, Innovation and Scholarships through the proposed collaborations on globally competitive research and innovation. These strategic international engagements will enhance the research and technological capacity for innovation within South Africa, as well as ensure the establishment and maintenance of research infrastructure and platforms. In this way, the project is also aligned to the Strategic Research Infrastructure Roadmap.

\section{Benefits}
The motivation for an optical transient network goes beyond purely the science when one considers applications of the concept in other fields. In its most distilled form, the success of the network depends on the ability to make {\bf sensible decisions} in {\bf real-time} using {\bf streamed data}. There are a host of other disciplines, such as medicine, security, early-warning systems etc, where such real-time decision-making can be employed to contribute to sustainable socioeconomic development. People involved in the BRICS-OTN will therefore acquire the same skills need in these other disciplines.

The program also provides an ideal environment for training students in skills such as optimisation, machine learning and the management and analysis of large datasets. This is a crucial juncture between astrophysics, engineering and computer science, and is valuable not only for the science returns but also for the skills that students and scientists will develop at this interface. These skills are far more broadly applicable than astrophysics, and form the keystone of the 4th Industrial Revolution (4IR), which will see a merging of the physical, biological \& computational realms.

The vast quantities of data, and particularly streamed data, provides an interesting test bed for rapid, real-time reactions. As such, there is strong potential for industrial partners to join the collaboration in efforts to collaborate on bespoke computational and data storage solutions.

To ensure that skills are transferred both between countries in the project, and between fields that may benefit from these techniques, we propose three specific approaches to human capacity development and outreach:
\begin{enumerate}
\item Training of students and early career researchers in the access to and analysis of network-generated data through workshops. In this respect, there is a strong potential for overlap between this BRICS-OTN proposal and the BRICSKA proposal on Big Data (PI: R. Taylor).  We envisage that both programs could plan for 1 workshop per year with 20$-$30 attendees per workshop, which could be rotated amongst the BRICS partners.
\item Interdisciplinary workshops both to learn from and share with industry and other academic fields that use streamed data for real-time decision-making, e.g health sciences, disaster management, environmental monitoring. These would likely occur every 2$-$ years, once the full program was implemented. 
\item Extending participation to school children and the general public through principles of open research. This means making data available, in a web-friendly format, through portals such as Zooniverse (www.zooniverse.org) or Open Universe (http://www.openuniverse.asi.it/, an initiative led through the partners in Brazil), and using such data both for education and engagement of the public in astronomical research (e.g. 0.5 FTE, shared between the BRICS-OTN and BRICSKA program for 5 years to integrate data with these platforms).
\end{enumerate}

One wider benefit for humanity from establishing BRICS-OTN is potential disaster mitigation, through one of the main science drivers to detect and charaterise the orbits of Potentially Hazardous Objects (PHOs) amongst the population of Near Earth Objects (NEOs). The 20 meter meteoroid that exploded over Chelyabinsk (Russia) on the 15 February 2013, at about 9:20 a.m. local time, was not detected by any early warning space- or ground-based observatories. More than 1600 people were injured in that event, although the situation could have been far worse (fatalities and damage to buildings and infrastructure) if it had been a more direct impact over a major city. 

One of the major lessons of that event was that to detect 10-m class NEOs we need a worldwide network of wide field telescopes to observe the entire sky at least every few hours. The BRICS-OTN will ensure detection of most of the dangerous celestial bodies (PHOs) with a possible accuracy to ensure a warning time of not less than 4 hours. This is a minimum time required for civil defense mitigation activity. The existing survey telescopes employed for NEO detection, such as the Catalina Sky Survey, Pan-STARRS, ATLAS and ZTF, don’t have enough a high enough survey rate to detect all of the PHOs. The proposed BRICS-OTN would help to solve this problem.\\

\section{New Facilities and Infrastructure}
In order to achieve the full aims of BRICS-OTN, namely to cover the \textit{entire} sky (i.e. 4$\pi$ steradians) to a depth of AB magnitude of at least 21 at a cadence of 0.5$-$1 hour will require the construction of several dozen new wide-field (20$-$25 square degree) telescopes. These will be dedicated to survey the sky to detect and followup transients. After the first phase of the program (networking the existing telescopes for transient followup), we will embark on the second phase of establishing a new global network of 1-m telescopes, to be located at existing observatories in the BRICS countries.

This is a long term program, expected to take ($\sim$10 years) to be fully realized develop, culminating in the  \textit{first dedicated} all-sky network of globally distributed 1-m class telescopes, all within the BRICS countries. This unique network would provide the enhanced capability of all-sky coverage at high cadence, as described in previous sections of this proposal. As previously stated, this future development builds mostly on the \textit{Sitian} concept, developed by the Chinese PI of this programme, Prof Jifeng Liu (NAOC), which will now form the basis of the proposed BRICS-OTN, though with some likely design modifications (see paper in these proceedings). 

\textit{Sitian} will consist of 72 (24 groups of 3) 1-m wide field (at least 5$^{\circ}$ $\times$ 5$^{\circ}$) telescopes which, when combined, will image the entire sky every $\sim$30 min, to a depth reaching 21.5 mag. Each field will be observed simultaneously in 3 filters  (\textit{g, r, i}), with 60-s exposures, with images being co-added to achieve the required depth. The telescope will be controlled by a central ``brain",  which will determine the optimal observing schedule, react to alerts and trigger follow-up observations, control each observation, handle the data management and processing. Future developments of \textit{Sitian} will include provision for spectroscopic followup, with dedicated 4-m class telescopes and expansion of the network globally.

To build the new optical transient network will require the construction of new telescope enclosures at the existing observatories in the BRICS countries (and other countries if there is interest to participate in the project). These building could be innovative in design, potentially housing 3 telescopes each.  

During the design review process for the new telescope builds within BRICS-OTN (to be conduced in the first year of the program), we will look at all the competing design options and conduct a technical capability cost$-$benefit trade-off study. In particular, we will consider the merits of the optical design of a 1-m telescope with a 7$^{\circ}$ $\times$ 7$^{\circ}$ FoV recently suggested by opticians from Russia. For both the design and construction phases, we will be almost certainly be considering a joint cooperative project between all BRICS participants for these new generation 1-m class super wide-field survey telescopes.

\section{Project Management \& Costing}
The governance and management structure will consist of:
\begin{itemize}
    \item Project Executive consisting of the Principal Investigators from each country. Overall responsibility for the project and budget authority. Report to BRICS and funding authorities.
    \item A Project Manager reporting to the Project Executive who would be responsible for day-to-day operations.
    \item Science  Working Groups with WG leads (at least two co-chairs from different participating countries).
      The WG leads coordinates activities in science areas, namely:
    \begin{itemize}
        \item radio transients; FRBs
        \item extreme cosmic explosions (GRBs, GW events/kilonovae)
        \item supernovae, TDEs, nuclear transients
        \item X-ray transients (Low and High Mass X-ray binaries)
        \item novae and binary star transients
        \item microlensing event, unusual stellar transients
        \item transient event followup from LSST and other alert streams
        \item Near Earth Objects, Potentially Hazardous Objects
        \item outreach, education and development
        \item networking software for telescopes for automated followup
        \item data processing pipelines
    \end{itemize}
\end{itemize}

\noindent The Project Executive plus Science Working Group leads will form a Project Management team. Annual project meetings will rotate amongst the participating countries. The program will also support smaller thematic workshops and hands-on data weeks.\\

In terms of the schedule of activities envisaged in BRICS-OTN, we adopt a phased approach. This starts with networking the existing facilities available within BRICS to allow for the automation of followup observations of transients from a variety of triggers, which will be a parallel development across several observatories within BRICS. As new funds become available, the program will evolve to the next stage of developing a network of new wide-field telescopes, ramping up over a 5$-$10 years. \\

\noindent The budget for the BRICS Optical Transient (BRICS-OTN) Network would consist of funds to support the following:

\begin{itemize}
    \item {\em BRICS-OTN annual project meeting}.  
    
    Cost for hosting an annual project meeting which will rotate among the member countries.
    
    \item {\em Travel for joint technical work, research collaboration meetings and workshops} 
    
    Travel for software engineers to work together for $\sim$weeks at a time at different partner institutions.
    
    Travel for co-investigators, postdocs and students to attend annual project meeting.
    
    Travel for extended visits by individual co-investigators to different partner institutions.
    
    Travel for small local workshops in science thematic areas.
    
    \item {\em BRICS post-doctoral and post-graduate fellowships}.  
    
    We propose six BRICS postdoctoral fellowships, covering the main science theme, plus twelve postgraduate PhD studenships.  These individuals will work on various transient science projects that exploit BRICS-OTN and the various supporting multi-wavelength facilities within the BRICS countries.  
    
    These positions will serve not only to further the transient astrophysics programmes, but to also reinforce science collaboration among the partners.  The postdoctoral fellowships will be joint appointments at institutions from at least two participating BRICS countries. The postgraduate students will be jointly supervised by co-investigators from at least two relevant BRICS partner countries.
    
    \item {\em Project management and software and system development}. 
    
    A BRICS-OTN Project Manager will be appointed at one of the participating BRICS institutes, probably the SAAO.
    
    We propose five technical appointments, namely software engineers, programmers, etc, for development of telescope networking and scheduling software systems.  One each would be appointed at relevant participating institutes from each BRICS partner country.

    A 25\% FTE appointment (possibly shared with the Big Data program) to support administrative and logistical needs of the project.
    
    \item {\em Outreach and Astronomy for Development Support}
    
    A 25\% FTE appointment to manage the outreach and development program (possibly shared with the Big Data program).

 \item {\em Equipment and Infrastructure}
    
    Equipment budget for postdoctoral and postgraduate fellows, including for laptop computers and associated hardware.
    
    The largest component will be for new control cards, firmware and software to enable existing telescope upgrades to allow them to be integrated within a local network.
    
    \item{\em SALT and other large telescope access costs}
    
    For spectroscopic followup of transient discoveries, the BRICS-OTN will budget for an annual cost to obtain 10\% of the observing time on SALT. 
    
    Should funding allow, we will also consider purchase of time on other telescopes in the 4-m to 10-m range.

    \item {\em Costs for new wide-field 1-m telescopes}
    
    Costs for designing and fabricating the 72 new telescopes that would make up the second phase of the BRICS-OT. These dedicated survey and followup telescopes would be co-located with exiting telescopes comprising the first phase of BRICS-OTN, at existing BRICS observatories.
    
    The budgets for new builds will be reviewed annually and de-scope options considered where needed.

\end{itemize}

For the first year of the project, 2020, we request funds for travel and for the BRICS-OTN project meeting and for appointing a Project Manager.  At that meeting we would implement the project governance and management structure, create detailed plans and agreements for programs and activities in the thematic areas, finalize projects for initial post-doctoral and post-graduate appointments.  Funds for the full projects, including fellowships and appointment of technical staff would be required beginning in the second year.

In addition, we would request funds for SALT access to allow collaborators within the BRICS-OTN to undertake followup observations of transients immediately. 

The estimated costs for each category, by year, are provided in budget table in an appendix of this paper, which total $\sim$ 280M\euro.  Amounts are all demarcated in the currency of Euros and represent the total costs of the activities. The BRICS Astronomy Working Group and the principals/governments of each BRICS country will need to negotiated how the costs are apportioned (e.g. by GDP fraction, user fraction, etc.). 

Budget de-scope options include the scaling down of the build for the 72 new telescopes beginning in phase 2 and 3 of the schedule. Clearly a reduction in the number of new telescope constructed will impact of the main science aims of the second phase of BRICS-OTN, namely the have the world's first all-sky patrol network for transients. However the full realisation of the program could still happen if the development was extended over further years so the spending profile could be accommodated by the available budget. Clearly there is also the possibility to lobby for funding within BRICS and worldwide to support this ambitious world-changing project which would firmly put BRICS on the map in terms of compelling science, technology and innovation.\\

\section{Summary}
In summary, based on the positive results already obtained with existing (smaller-scale) monitoring projects, and driven by the goal of high cadence all-sky observations, we propose a much more ambitious, ground-breaking and globally  compelling project: an automated global network of telescopes for transient followup and detection. This will initially include harnessing existing facilities in all of the BRICS countries, followed by the establishment of a distributed network of many ($\sim$70) wide field 1-m optical telescopes. This new network will map the whole sky every hour or so and truly open a new time-domain window on the universe with unprecedented sky coverage and cadence. This proposed \textit{BRICS Optical Transient Network (BRICS-OTN)} will allow us to expand our transient ``discovery space" into regions hitherto unexplored, even for the Rubin Observatory's Legacy Survey of Space and Time (LSST). Continuous access to the whole sky with such wide field telescopes, equipped with fast detectors (e.g. sCMOS devices), will allow for both high time resolution and continuous monitoring observations of transients objects, which could also support multi-wavelength and multi-messenger observations. The programme will leverage state of the art Big Data/Big Compute facilities and will build on the strength and opportunities that the BRICS consortium has developed in a collective and collaborative manner.  Importantly it will provide unique opportunities for young researchers in all of the BRICS countries to lead in a rapidly evolving and scientifically leading field of astronomy in the 21st century.  Human capacity development and other spin-off benefits will greatly benefit the individual BRICS countries, through both the tangible outcomes, like the training new cohorts of scientists and engineers, but also through the less tangible benefits which align with sustainable developments goals and empowerment. 

\section{Acknowledgments}\label{acknowledgments}
The authors are grateful to the many (over 60) co-investigators, many of whom assisted in developing this BRICS proposal. DAHB thanks the South African Department of Science and Innovation for support to attend the BRICS Astronomy Group (BAWG) meeting in Rio de Janeiro.  We also thank the local Brazilian hosts, particularly Bruno Vaz Castilho de Souza, Ulisses Barres de Almeida and Carlos Alexandre Wuensche.

\bibliographystyle{abbrv}
\bibliography{transients-bawg-2019-final.bib}

\newpage
\pagebreak
\begin{sidewaystable}
	\caption{Cost Estimates for BRICS-OTN}	

\begin{center}
    \begin{tabular}{|l|r|r|r|r|r|r|r|r|r|r}
    \hline
    \multicolumn{10}{|c|}{{\bf Costs ('000 Euros)}} \\
    \hline
    {\bf  Year} & {\bf 2020} & {\bf 2021} & {\bf 2022} & {\bf 2023} & {\bf 2024} & {\bf 2025} & {\bf 2026}& {\bf 2027} & {\bf 2028} \\
    \hline
       Host Annual Meeting & 6 & 6 & 7 & 7 & 7 & 8 & 8 & 10 & 10\\
       \hline
       Travel & & & & & & & & & \\
       \quad Annual Meeting & 85 & 121 & 127 & 134 & 140 & 147 & 155 & 160 & 166 \\
       \quad Visits/workshops &&25 & 26 & 28 & 29 & 30 & 31 & 32 & 34\\
       
       HCD &&&&&&&&&\\
         \quad Postdoc Fellows &&112 &118 & 124 & 130 & 136  & 143 & 146 & 150\\
         \quad Postgrad Students && 105 & 110 & 115 & 121 & 127 & 134 & 136 & 138\\
       \hline
       Staff &&&&&&&&&\\
       \quad Project Manager &23 & 47 & 49 & 51 & 54 & 57 & 59 & 62 & 65\\
       \quad Software Engineers &&232& 244 & 256 & 269 & & & & \\
       \quad Admin \& Logistics &6 & 11 & 12 & 13 & 13 & 14 & 15 & 16 & 18\\
       \hline
       Outreach/development & & 23& 24 & 25 & 27& 28 & 30 & 33 & 36\\
       \hline
       Computing resources & &26 & & 60&& 70& & & 90\\
       \hline
        Networking hardware & &22 & 35 & 50 &&&&& \\
        \hline
       SALT access (10 \%) & 330 & 360 & 400 & 440 & 480 & 520 & 560 & 600 & 650\\
       \hline 
       New telescopes (72) & & & 10,000 & 20,000 & 40,000 & 50,000 & 50,000 & 50,000 & 50,000\\
       \hline 
       {\bf Total} & 450 &1,084 & 11,152& 21,303 & 41,263 & 51,137 & 51,135 & 51,195&  51,357\\
       \hline
    \end{tabular}
    \label{tab:budget}
\end{center}
\end{sidewaystable}

\end{document}